%% file: g8_KY_Pol_Obs.tex
\documentclass[%
 reprint,
superscriptaddress,
 amsmath,amssymb,
 aps,
]{revtex4-2}

\usepackage{graphicx}
\usepackage{dcolumn}
\usepackage{bm}
\usepackage{amsmath}
\usepackage{mathrsfs}
\usepackage{float}

\usepackage{color}
\newcommand{\obs}[1]{\mathbf{#1}}

\usepackage{orcidlink}
\usepackage{soul}

\begin{document}

\title{Photoproduction of the $\Sigma^+$ hyperon using linearly polarized photons with CLAS}


\input{authors.tex}


\date{\today}
\begin{abstract}
\begin{description}
\item[Background]
Measurements of the polarization observables $\obs{\Sigma}, \obs{P}, \obs{T}, \obs{O_{x}}, \obs{O_{z}}$ for the reaction $\vec{\gamma} p \rightarrow K_{S}^0 \Sigma^{+}$ using a linearly polarized photon beam of energy 1.1 to 2.1~GeV are reported.
\item[Purpose]
The measured data provide information on a channel that has not been studied extensively, but is required for a full coupled-channel analysis in the nucleon resonance region.
\item[Method]
Observables have been simultaneously extracted using likelihood sampling with a Markov-Chain Monte-Carlo process. 
\item[Results]
Angular distributions in bins of photon energy $E_{\gamma}$ are produced for each polarization observable.
$\obs{T}, \obs{O_{x}}$ and $\obs{O_{z}}$ are first time measurements of these observables in this reaction. The extraction of $\obs{\Sigma}$ extends the energy range beyond a previous measurement. The measurement of $\obs{P}$, the recoil polarization, is consistent with previous measurements.
\item[Conclusions]
The measured data are shown to be significant enough to affect the estimation of the nucleon resonance parameters when fitted within a coupled-channels model.

\end{description}
\end{abstract}

\pacs{??, ??, ??, ??, ??, ??}

\maketitle


\section{\label{sec:introduction}Introduction}

The database of observables from photoproduction of mesons in the nucleon resonance region now numbers tens of thousands of points \cite{thiel_light_2022, ireland_photoproduction_2020}. Together with a similar number of pion scattering data points, they represent the experimentally measured information that is available to interpret the spectrum of light baryons, i.e. determine which resonances exist and some of their properties. 

Most of the data is in the form of differential cross sections, but a substantial amount of polarization observables have been measured, thanks to advances in the ability to produce polarized beams of photons (linear and circular polarization), polarized targets (longitudinal and transversely), and to measure recoil polarization. 

In the hadronic mass region of roughly 1 to 3~GeV, many meson-baryon final states can be produced, and it is now generally accepted that a sophisticated theoretical model requires a full coupling of all channels to be able to determine the light baryon spectrum. Hence the need to produce a comprehensive dataset. Whilst a range of different final state reactions have been measured ($\pi N$, $\eta N$, $KY$, vector mesons, multiple mesons), there remain significant differences in the amount and quality of data available for each channel.

Strangeness photoproduction is seen as ideal for determining many polarization observables, due to the weak decay of hyperons giving rise to self-analysing final state recoiling baryons. In principle it is now possible to carry out measurements that can determine a sufficient number of different observables to extract amplitude information \cite{Sandorfi_2011}. However, amplitude extraction also requires these measurements to be performed to sufficient accuracy \cite{ireland_information_2010}. The usefulness of any dataset is generally determined by the sensitivity of fits of theoretical models to the measured data. 

In this paper, we report a measurement of five polarization observables for the reaction 
$\vec{\gamma} p \rightarrow K_{S}^0 \Sigma^{+}$ using a linearly polarized photon beam of energy 1.1 to 2.1~GeV (invariant energy W = 1.716 to 2.195~GeV). This channel has been previously measured to give values for recoil polarization $\obs{P}$ and photon beam asymmetry $\obs{\Sigma}$. Our present measurement of $\obs{P}$ overlaps the previous data \cite{nepali_2013,EWALD2014268} which allows a check for consistency. It extends the range of measured $\obs{\Sigma}$, and is the first measurement of the target spin asymmetry $\obs{T}$, and the beam-recoil double-spin asymmetries $\obs{O_x}$ and $\obs{O_z}$. 


\section{\label{sec:experiment}Experimental Setup and Formalism}

The Thomas Jefferson National Accelerator Facility (JLab) in Newport News, Virginia is home to the Continuous Electron Beam Accelerator Facility (CEBAF). The data presented here were acquired in 2001, during the g8b experiment, utilising the delivered electron beam and both the CEBAF Large Acceptance Spectrometer (CLAS) detector \cite{MECKING2003513} and the photon tagging spectrometer (tagger)\cite{sober_bremsstrahlung_2000} in Hall B. Linearly polarized photons were produced by scattering electrons from a diamond radiator, a technique known as coherent bremsstrahlung \cite{timm_coherent_1969,lohmann_linearly_1994}, with the energy-degraded electron detected in the photon tagging spectrometer. 
Final state particles were detected in the CLAS detector, which was a multi-subsystem detector package utilising drift chamber (DC) tracking, time-of-flight (ToF), calorimetry and in the case of photoproduction experiments, a start time counter (ST) arranged in six sectors \cite{Mestayer:2000we, Smith:1999ii, Amarian:2001zs, taylor2001clas}. In combination with a toroidal magnetic field, these subsystems allowed for the reconstruction of both charged and neutral particles with momentum resolution $\sigma_p/p \sim $1\%  and a polar angle acceptance from 8$^{\circ}$ to 140$^{\circ}$. The target used in this experiment was a 40-cm-long liquid-hydrogen target and was placed 20\,cm upstream from the geometric centre of CLAS. 
The experimental configuration used for this data collection consisted of a 4.55~GeV electron beam incident on a 50-$\mu$m thick diamond radiator, producing a beam of linearly polarized photons. Five coherent edge settings, where the degree of linear polarization is maximal, were selected \cite{livingston_stonehenge_2009} at 1.3, 1.5, 1.7, 1.9 and 2.1~GeV, and the degree of polarization was calculated analytically from a fit \cite{livingston_polarization_2011}. The data were collected in two orthogonal orientations of the photon linear polarization, parallel and perpendicular to the laboratory floor, allowing systematic uncertainties relating to detector acceptance to be minimised.

\begin{figure}[h]
  \centering
  \includegraphics[width=0.9\columnwidth]{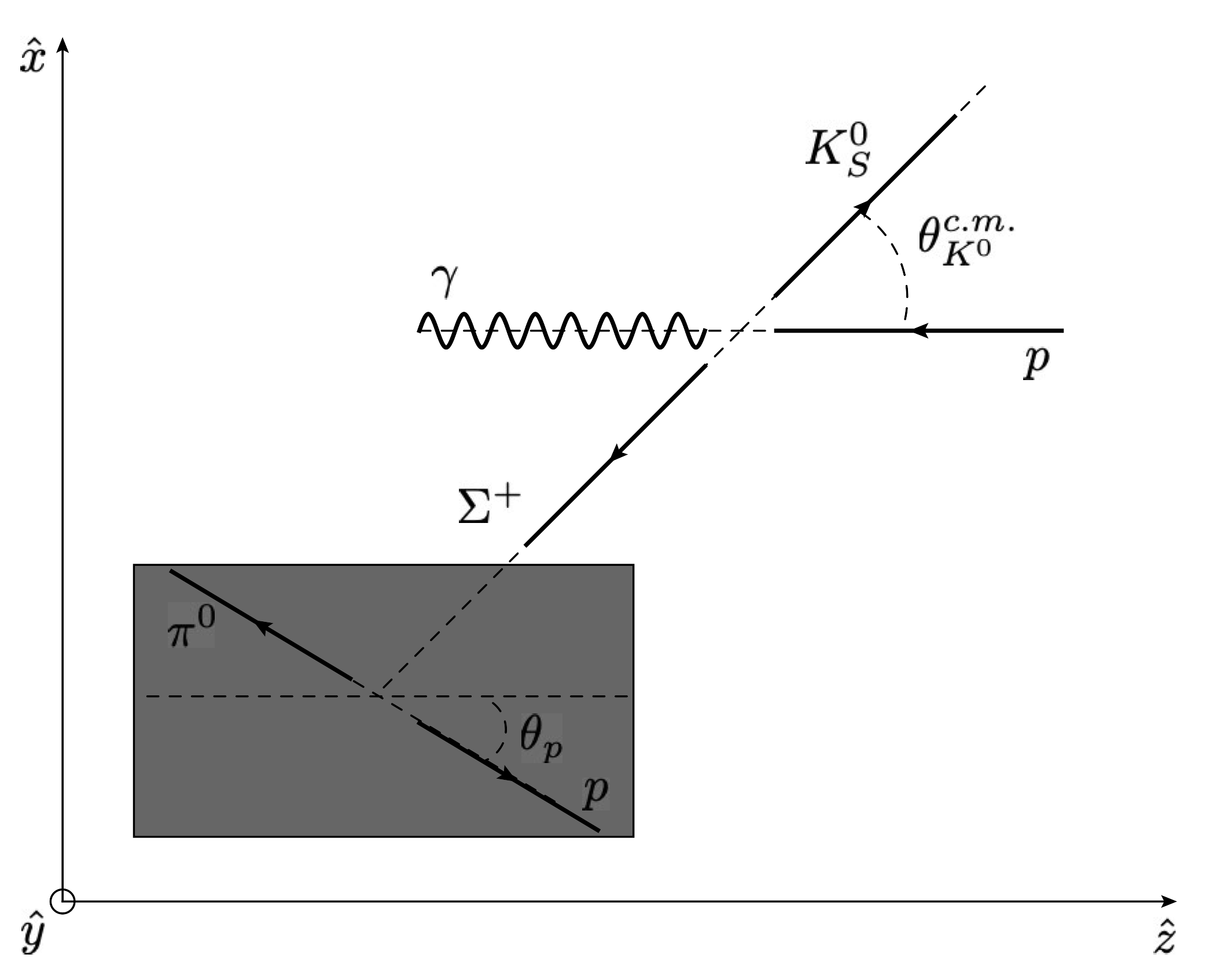}
  \caption{Kinematics of the $\gamma p\rightarrow K^{0}\Sigma^+$
reaction.}
\label{fig:Kinematics}
\end{figure}

The coordinate system and kinematic variables used in the description of kaon photoproduction are shown for the centre-of-mass reference
frame in Figure \ref{fig:Kinematics}.

The unprimed coordinate system defined in the rest frame of the hyperon is chosen for this analysis, where the z-axis is orientated along the momentum of the boosted incoming photon ($\vec{k}$). With a kaon momentum $\vec{q}$, $\hat{x}$, $\hat{y}$ and $\hat{z}$ are defined as: 
\begin{equation}
\hat{z}=\frac{\vec{k}}{\left|\vec{k}\right|};\quad\hat{y}=\frac{\vec{k}\times\vec{q}}{\left|\vec{k}\times\vec{q}\right|};\quad\hat{x}=\hat{y}\times\hat{z}\label{eq:unitUnprimed}.
\end{equation}


\section{\label{sec:events}Event Selection}

In this measurement, neither the $K^{0}_{S}$ nor the $\Sigma^{+}$ were directly detected. Instead the reaction was determined and isolated via the identification of charged final state particles:
\[
\overrightarrow{\gamma}p\rightarrow K^{0}_{S}\Sigma^{+}\rightarrow p\pi^{+}\pi^{-}\left(\pi^{0}\right),
\]
where the $K^{0}_{S}$ decays into a $\pi^{+}$ and $\pi^{-}${\huge{}} with a 69.3\% branching ratio, whilst the $\Sigma^{+}$ decays into a proton and $\pi^{0}$ with a 51.6\% branching ratio. The $\pi^{0}$ was reconstructed from the $p\pi^{+}\pi^{-}$ missing mass (MM), while the $\Sigma^{+}$ and $K^{0}_{S}$ were subsequently reconstructed from the invariant mass (IM) of $p\pi^{0}$ and $\pi^{+}\pi^{-}$, respectively.

The analysis was carried out for each of the five coherent peak positions, with an initial cut applied to restrict the data to the photon energies with a relatively high degree of polarization. The polarized photons for each setting were identified in a 200-MeV-wide bin with an upper limit at the coherent edge position. The 200-MeV bin size was found to be optimal for consistency of the polarization value, as demonstrated by the study described in \cite{paterson_photoproduction_2016}.

\subsection{Event Filter and Particle Identification (PID)}

After an initial selection of events with the required number of particles as well as a valid hit in the tagging spectrometer, a loose identification of the particles was made using the charge and mass determined by the drift chambers (DC) and time-of-flight (ToF) subsystem, and a z-vertex coordinate compatible with the target geometry. Based on these requirements and considering the final state particles that would arise from the decays of the $K^{0}_{S}$ and $\Sigma^{+}$, events with exactly one proton, $\pi^{+}$, and $\pi^{-}$ (plus possible neutrals) were retained. Although not explicitly required for the reaction of interest, the possibility of a neutral being in the data is retained.
Having determined the three final state hadrons, the remaining particle identification step was to associate them with the correct detected photon. The hadron timing information was extrapolated backwards to the event vertex. This vertex time should be identical to the tagger timing of the event photon, within the 2~ns beam bucket delivered by CEBAF. The interaction photon was identified based upon this time coincidence.

\subsection{Channel Identification}

After the application of the particle identification procedure described, the following mass requirements were placed on the reconstructed particle combinations:

\begin{itemize}
    \item $\pi^{0}$: 0.05 $<$ MM($p \pi^{+} \pi^{-}$) $<$ 0.22~GeV/c$^{2}$
    \item $K^{0}$: 0.450 $<$ IM($\pi^{+} \pi^{-}$) $<$ 0.550~GeV/c$^{2}$
    \item $\Sigma^{+}$: 1.150 $<$ MM($\pi^{+} \pi^{-}$) $<$ 1.250~GeV/c$^{2}$.
\end{itemize}

\begin{figure}[h]
\begin{centering}
\includegraphics[width=0.9\columnwidth]{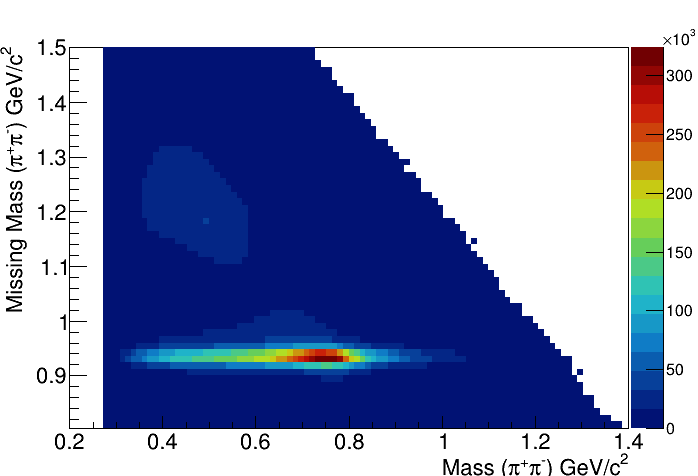}
\includegraphics[width=0.9\columnwidth]{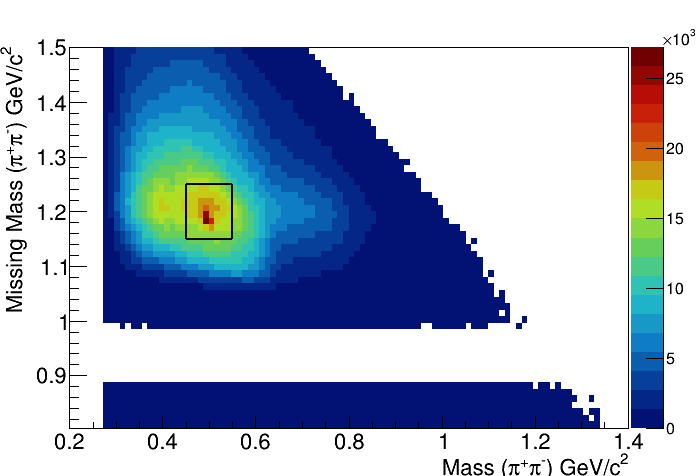}
\par\end{centering}
\caption{\label{fig:invmass}Missing mass ($\pi^{+} \pi^{-}$) vs. mass ($\pi^{+} \pi^{-}$). Top: after PID only. Bottom: after $\pi^{0}$ mass cut. The rectangle shows the extent of the mass cuts around the kaon and hyperon, and contains those events retained for further analysis.}
\end{figure}

Events remaining after the above cuts were used as candidates for the $K^{0}_{S}\Sigma^{+}$ final state. Distributions for the different mass combinations are shown in Figure \ref{fig:invmass}, where actual signal events are evident as described in the caption. The particle and channel identification cuts applied, and the number of events after each stage, are summarised in Table \ref{tab:Analysis-cuts}

\begin{table*}
\caption{\label{tab:Analysis-cuts} Analysis cuts applied and resulting number of events for combined coherent peak settings.}
\begin{ruledtabular}
\begin{tabular}{p{5cm}p{7cm}r}
Applied Cut & 
Details &  
Events \\
\colrule
Initial skim                                                                       & 
3 charged particles, optional neutrals in final state                              & 
$6.04 \times 10^{8}$  \\
z-vertex cut, proton and pion mass cuts                                            &
$-40<z<0$ cm   \newline 0.49 $<$ $M^{2}(p)$ $<$ 1.44~GeV/c$^{2}$  \newline  $M^{2}(\pi^{+/-})$ $<$ 0.09~GeV/c$^{2}$                                                  & 
$2.44 \times 10^{8}$   \\                                           
$\gamma$-proton vertex timing                                                      & 
$|$Vertex time($\gamma$) - Vertex time(p)$|$ $<$ 1.0~ns                            & 
$1.48 \times 10^{8}$ \\
Polarization                                                                       & 
Entry exists in polarization tables                                               & 
$5.06 \times 10^{7}$ \\
Mass cuts                                                                          & 
$\pi^{0}$: (GeV/c$^{2}$) $\in$ (0.05,0.22) \newline
$K^{0}$: (GeV/c$^{2}$) $\in$ (0.45,0.55) \newline
$\Sigma^{+}$: (GeV/c$^{2}$) $\in$ (1.15,1.25)                                      & 
$1.15 \times 10^{6}$
\end{tabular}
\end{ruledtabular}
\end{table*}

\subsection{\label{sub:photon-beam} Photon Beam Polarization}

The coherent bremsstrahlung process \cite{timm_coherent_1969,lohmann_linearly_1994}, in which the electron scatters from a diamond radiator producing linearly polarized photons, is well understood. The coherent enhancement above the $\sim1/E_{\gamma}$ incoherent unpolarized background distribution was measured and fitted using the method described in Reference \cite{livingston_polarization_2011} to calibrate the degree of linear polarization.  The range of beam polarization was 50 – 90\% in this experiment.
Systematic uncertainties in the degree of polarization are consistent for all reactions. Therefore we used a detailed study of the consistency of the calculated polarization using the high-statistics reaction $\gamma p\rightarrow p\pi^{0}$ \cite{dugger_beam_2013} \cite{dugger_note2011} to quantify this uncertainty. From this study the estimated systematic uncertainty of the photon polarization was 4\% for the 1.3, 1.5, 1.7, 1.9~GeV settings and 6\% for the 2.1~GeV setting.

\subsection{\label{sub:Signal-Background} Signal/Background Separation}

Prior to analysing the angular distributions, which are sensitive to the observables of interest, background contributions were removed. In this work, the sPlots technique, developed by Pivk and Diberder \cite{pivk2005}, was employed.

\begin{figure}[h!]
  \centering
     \includegraphics[width=0.9\columnwidth]  {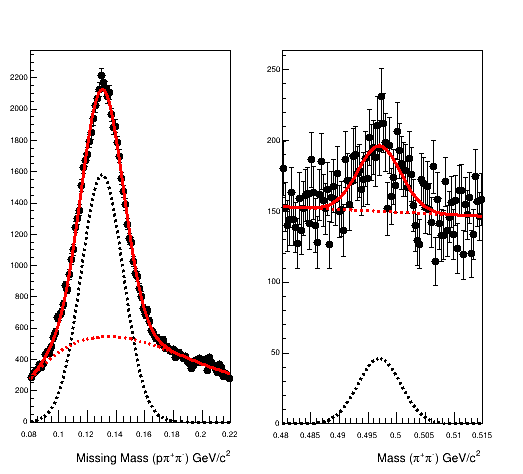}
  \caption{Example $\pi^{0}$ (left) and $K^{0}$ (right) mass fits for the $E_{\gamma}$ = 1.23~GeV and $\cos{\theta^{c.m.}_{K^0}}$ = -0.28 bin. In each, the plot shows data points in black, combined signal and background model in solid red, signal in dotted black, and background in dotted red.}
\label{fig:splot}
\end{figure}

To separate exclusive $p \pi^{+} \pi^{-} \pi^{0}$ events the missing mass of $p \pi^{+} \pi^{-}$ was used, which is peaked at the $\pi^{0}$ mass  for exclusive signal events. To then filter the strange $K^0\Sigma^+$ events the invariant mass of $\pi^{+}\pi^{-}$ was used, which peaks at the mass of $K^0_{S}$, above a background of exclusive, non-strange  $p \pi^{+} \pi^{-} \pi^{0}$ events. Weights were first obtained from the $\pi^{0}$ fit and used to form a weighted $K^0_{S}$ mass distribution.  A fit to this second distribution  provided a combined set of weights to extract exclusive $K^0_{S}\Sigma^+$ events, which were then used in extracting the polarization observables. In each case, the distribution was modelled as a Gaussian peak on a Chebyshev polynomial background. Figure \ref{fig:splot} shows an example of each of these fits for one bin in $E_{\gamma}$ and $\cos{\theta^{c.m.}_{K^0}}$.

\begin{figure}[h]
  \centering
  \includegraphics[width=0.9\columnwidth]{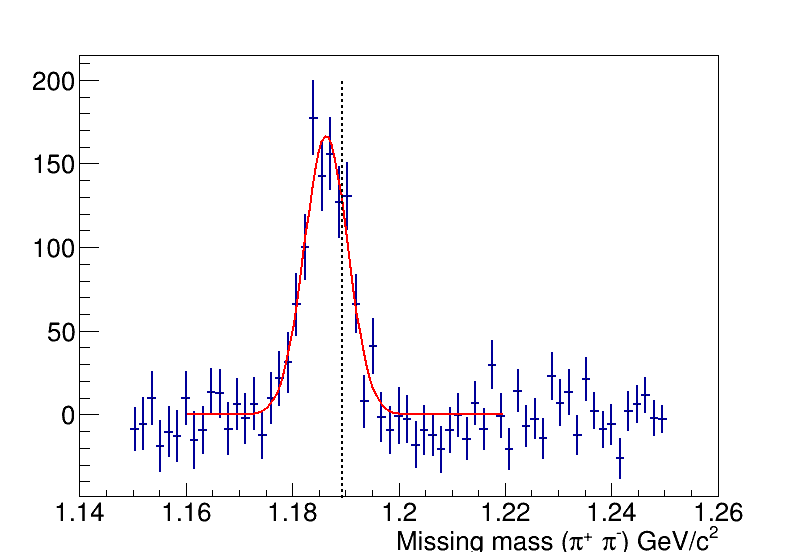}
  \caption{Missing mass ($\pi^{+}, \pi^{-}$) for one $\cos{\theta^{c.m.}_{K^0}}$ and $E_{\gamma}$ bin. The dashed vertical line shows the PDG mass of the $\Sigma^+$.}
\label{fig:sigmaMass}
\end{figure}

As a validation of the background subtraction procedure we show the subtracted missing mass of $\pi^{+} \pi^{-}$, which, for the case of exclusive signal events only, should give a narrow peak at the $\Sigma^0$ mass of 1.189~GeV$/c^2$. Figure \ref{fig:sigmaMass} shows an example of this mass distribution for one $\cos{\theta^{c.m.}_{K^0}}$ bin at $E_{\gamma}$ = 1.23~GeV.  The fitted mean of this $\Sigma^{+}$ mass varies from 1.185 to 1.188~GeV$/c^2$, within 0.3\% of the PDG value, across all bins.

\section{\label{sec:observables}Extraction of Observables}

This work reports one of several reactions measured in the same run period, and follows on from the results of Paterson {\it et al.}~\cite{paterson_photoproduction_2016}. In that work, a method was introduced for simultaneously determining all polarization observables to which the experiment was sensitive, using event-by-event maximum likelihood weights. There are several developments in the extraction technique since the work of Reference \cite{paterson_photoproduction_2016}: 
\begin{itemize}
    \item There is no independent measurement of $\Sigma^{+}$ recoil polarization $\obs{P}$, so it is not solely an asymmetry measurement in photon polarization, and an acceptance calculation must be included.
    \item The likelihood as a function of the observables of interest is sampled with a Markov-Chain Monte Carlo (MCMC)~\cite{Bonamente2017} to estimate the values and co-variances via a Metropolis-Hastings algorithm~\cite{Metropolis:1953am,HASTINGS1970}.
    \item To take into account the correlations among the observables, which arise from the spin algebra of pseudoscalar meson photoproduction, the MCMC samples amplitude space. The mapping from amplitudes to observables is surjective (many to one), so this would only be possible with a sampling algorithm, not an optimizer.
\end{itemize}

In order to construct a likelihood function, one must provide a probabilistic
model for what the data would be, given the model with its parameters
set to specific values. The differential cross section for the reaction,
assuming linear photon polarization only and that recoil polarization
can be determined, can be written \cite{thiel_light_2022}:
\begin{align}
\frac{d\sigma}{d\Omega} &\equiv
\sigma\left(\phi,\cos\theta_{x},\cos\theta_{y},\cos\theta_{z}\right) 
\nonumber \\
&= \sigma_{0}\left\{ 1-P^{\gamma}\obs{\Sigma}\cos2\phi\right.
\nonumber \\
&\quad -\alpha\cos\theta_{x}P^{\gamma}\obs{O_{x}}\sin2\phi
\nonumber \\
&\quad +\alpha\cos\theta_{y}\obs{P}-\alpha\cos\theta_{y}P^{\gamma}\obs{T}\cos2\phi
\nonumber \\
&\quad \left.-\alpha\cos\theta_{z}P^{\gamma}\obs{O_{z}}\sin2\phi\right\},
\end{align}\label{eq:beam-recoil-1}where $\sigma_{0}$ represents the unpolarized cross section, $P^{\gamma}$
is the degree of linear photon polarization, $\phi$ is the centre-of-mass azimuthal angle from the photon polarization plane to the kaon transverse momentum direction, the variables $\cos\theta_{x},\cos\theta_{y},\cos\theta_{z}$ are
the direction cosines of the decay proton in the $\Sigma^{+}$ rest frame and
$\alpha$ is the weak decay parameter, with value $-0.982\pm0.014$ \cite{workman_pdg}, for the reaction $\Sigma^+ \rightarrow p \pi^0$. The emboldened symbols represent the polarization observables that we wish to extract. 

We can thus construct a function $\mathscr{I}\,(=$$\,\frac{d\sigma}{d\Omega})$ that depends on both the measured variables 
$\mathbf{x} =\{ P^{\gamma}, \phi, \theta_x, \theta_y, \theta_z\}$
and the parameters we wish to estimate (polarization observables), $\mathbf{\Theta} =\{\obs{\Sigma}, \obs{O_x}, \obs{O_z}, \obs{T}, \obs{P}\}$. An acceptance function $\eta(\mathbf{x})$, determined by simulation is also required, so if each event is characterised by its ``data'' $\mathbf{x}_i$, we have for the likelihood of the N event sample:
\begin{align}\label{eq:like1}
\mathscr{L}\left(\mathbf{\Theta}\right) = \displaystyle\prod_{i}^{N} \frac{\mathscr{I}\left(\mathbf{x}_i:\mathbf{\Theta}\right)\eta\left(\mathbf{x}_i\right)}{A\left(\mathbf{\Theta}\right)}.
\end{align}
$A\left(\mathbf{\Theta}\right)$ is the probability normalisation integral, integrated over the full range of the measured observables, given by
\begin{align}\label{eq:like2}
A\left(\mathbf{\Theta}\right) = \int \mathscr{I}\left(\mathbf{x}:\mathbf{\Theta}\right)\eta\left(\mathbf{x}\right) \,d\mathbf{x}.
\end{align}

The sPlot process gives a weight $w_i$ for each event. The inclusion of weights in the likelihood means that an additional factor is required to account for the effect of the weights on the uncertainties. Specifically, the uncertainty will scale with the quantity $\frac{\sum_i w^2_i}{\sum_i w_i}$ and we apply an approximate correction factor accounting for this in the log likelihood,  which is given by:

\begin{align}\label{eq:like4DG}
\ln \mathscr{L}\left(\mathbf{\Theta}\right) =&\frac{\sum_i w_i}{\sum_i w^2_i}\displaystyle\sum_{i}^{N} 
\left[w_i \ln \frac{ \mathscr{I}\left(\mathbf{x}_i:\mathbf{\Theta}\right)\eta\left(\mathbf{x}_i\right)}{A\left(\mathbf{\Theta}\right)} \right]\\
 =& \frac{\sum_i w_i}{\sum_i w^2_i}\displaystyle\sum_{i}^{N} 
\left[w_i \ln \frac{ \mathscr{I}\left(\mathbf{x}_i:\mathbf{\Theta}\right)}{A\left(\mathbf{\Theta}\right)} + \ln\eta\left(\mathbf{x}_i\right) \right],
\end{align}
where we ignored the $\ln\eta\left(\mathbf{x}_i\right)$ term in the sampling as it does not effect the parameter posterior distributions, and we approximated   $A\left(\mathbf{\Theta}\right)$ by

\begin{align}\label{eq:like2approx}
A\left(\mathbf{\Theta}\right) = \sum^{N_{s}}_{s} \mathscr{I}\left(\mathbf{x}_s:\mathbf{\Theta}\right),
\end{align}
where the sum over $s$ is over $accepted$ Monte-Carlo events in which the full reaction has been simulated (using a phase space generator \cite{edgen_2017} and weighted to reflect the distributions in $E_{\gamma}$ and four-momentum transfer, $t$, observed in the data), reconstructed and passed through our selection cuts. This accounts for the factor $\eta\left(\mathbf{x}_s\right)$ in the integral. 

The task of the MCMC is to draw samples from a distribution that will asymptotically match the distribution, so that the samples can be used to determine estimators of the quantities of interest. We are therefore looking for a distribution in the parameter space $\Theta$. 

However, there is a further restriction. The five observables that this measurement is sensitive to are part of a full set of 16 observables that are related to four (complex) amplitudes, which are functions of $W$ and $\cos{\theta^{c.m.}_{K^0}}$. The mapping from amplitudes to observables depends on the basis in which the amplitudes are defined. Using the transversity basis, such as that proposed in Vrancx {\it et al}.\cite{vrancx2013}, the four complex transversity amplitudes $a_j$ are normalised such that 
\begin{equation}
    a_1^2 + a_2^2 + a_3^2 + a_4^2= 1,
\end{equation}
where $a_j^2 = a_j a_j^{\star}$, and which divides out an overall factor proportional to the cross section. The real and imaginary components of the amplitudes thus form a unit hypersphere in 8-dimensional space (7-sphere). The polarization observables are then related to the normalised amplitudes by \cite{vrancx2013}:

\begin{align}
\Sigma &= a_1^2 + a_2^2 - a_3^2 - a_4^2
\nonumber \\
P &= a_1^2 - a_2^2 + a_3^2 - a_4^2
\nonumber \\
T &= a_1^2 - a_2^2 - a_3^2 + a_4^2
\nonumber \\
O_x &= 2 \operatorname{Re} \left(a_1 a_4^* + a_2 a_3^* \right)
\nonumber \\
O_z &= 2 \operatorname{Im} \left(a_1 a_4^* - a_2 a_3^* \right).
\end{align}\label{eq:amp-to-obs}

For the MCMC used here, the sampling proceeds in the parameter space of the real and imaginary parts of amplitude space, where it is straightforward to explore the surface of the 7-sphere. The proposed values ${a_j}$ are then used to evaluate proposed values of $\Theta$ via Equation 4, at which the likelihood is evaluated. The MCMC sampling was performed sequentially by drawing a randomized step size for a single random amplitude part and moving to that point. The Metropolis-Hasting algorithm~\cite{Metropolis:1953am,HASTINGS1970} was then followed until 2000 steps had been accepted in the chain. This number included 50 burn-in steps that were excluded from the posterior distributions.  

An example of the resulting sampled probability density functions is shown in Figure \ref{fig:Example-of-corner}. In this example, it can be seen that the MCMC amplitude sampling procedure ensures the limits of -1 and +1 (e.g.~$P$ and $T$), and correlations between different observables are evident (e.g.~$\Sigma$ vs.~$T$).

\noindent \begin{center}
\begin{figure}[h]
\noindent \centering{}\includegraphics[width=1.0\linewidth]{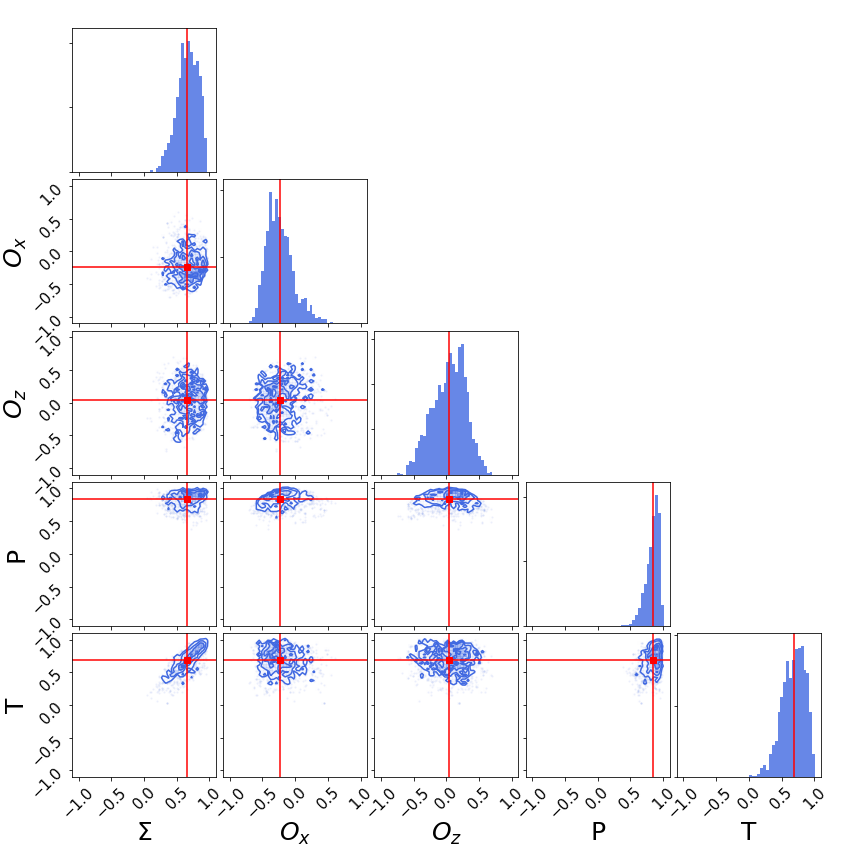}\caption{\label{fig:Example-of-corner}Example of a so-called corner plot, showing the posterior distributions of the observables evaluated at $W=1.786$~GeV and $\cos{\theta^{c.m.}_{K^0}}$=-0.28, together with the correlation plots.}
\end{figure}
\par\end{center}


\section{\label{sec:systematics}Systematic Uncertainties and Validation}

Systematic studies were performed by varying the width of the fit around the $\pi^0$ mass, the background shape and width of the fit around the $K^0$ mass and the MCMC step size. The dominant uncertainty, from the fit to the $K^0$ mass, was studied by obtaining signal weights when the interval was 35, 50, 65 and 80~MeV. Using each set of weights, MCMC chains were produced and merged to create a posterior distribution (PDF) that includes the effects of these systematic variations. The resulting effect on the extracted observable distribution variances was small compared to the statistical uncertainty.
In order to validate the extraction method, toy data was created from a phase space simulation with known fixed values set for each observable based upon the real data. Each event in the toy data was processed as accepted or rejected based on the probability of it being included in a distribution with the given observable values. The distributions of deviation, measured minus true, for each observable resulted in mean values of the order 0.01 to 0.02. These deviations are an order of magnitude smaller than the standard deviations quoted leading to the conclusion that any bias introduced by the fitting method is insignificant in comparison to the statistical uncertainty.
These systematic uncertainties from the fit interval and method validation were considered along with those from the photon polarization (see Section \ref{sub:photon-beam}) and the 1.7\% introduced from the $\Sigma^+$ weak decay parameter $\alpha$, resulting in an upper limit on the systematic uncertainty of 6\% for photon energy bins in the range 1.1 to 1.85~GeV and 7\% for the 1.85 to 2.1~GeV bin.


\section{\label{sec:results}Results}

Data for the five observables, determined at a total of 21 points in energy and angle, are presented in Figure ~\ref{fig:CompJuBo}. Plotted for comparison are the calculations of the Juelich-Bonn dynamical coupled-channels model ~\cite{Ronchen:2012eg} before (brown) and after (blue) a refitting of the model with the current data included. 

In the Juelich-Bonn model a hadronic potential derived from an effective Lagrangian with chiral constraints is iterated in a Lippmann-Schwinger equation formulated in time-ordered perturbation theory. The photo-interaction is described in a semi-phenomenological framework~\cite{Ronchen:2014cna}. The model preserves unitarity and analyticity and resonance states are defined as poles in the complex energy plane of the second Riemann sheet of the scattering amplitude. The values of the free parameters of the model are determined in simultaneous fits to pion- and photon-induced reactions. 

In its most recent version ``JuBo2022"~\cite{Ronchen:2022hqk} about 72,000 data points from the reactions $\pi N$, $\gamma p\to$ $\pi N$, $\eta N$, $K\Lambda$ and $K\Sigma$ were considered. Only 448 of those data points   came from the $\gamma p\to K^0\Sigma^+$ channel, compared to, e.g., 5632 for $\gamma p\to K^+\Sigma^0$. In combination with the fact that only very few polarization data were available, it was noticed in Ref.~\cite{Ronchen:2022hqk} that the determination of the $\gamma p\to K^0\Sigma^+$ amplitude is difficult and achieving a good description of the data is hard, resulting in a rather large total reduced $\chi^2$ of 3.16 for $K^0\Sigma^+$ compared to 1.66 for $K^+\Sigma^0$. 

In the re-fit including the new polarization data (``JuBo2023-1" solution), however, the description of older $K^0\Sigma^+$ data could also be improved and a total reduced $\chi^2$ of 2.01 for $K^0\Sigma^+$ was achieved. 
Whilst it is possible to see that there is some improvement in the model fit for all the observables, the improvement in the goodness of fit can be estimated by examining the reduced $\chi^2$ statistic. As can be seen from Table \ref{tab:fitChiSq}, there is indeed a noticeable improvement, indicating that the new data are able to affect the models. 

We examined the consequences of including the new data in the fit, and as shown in Table \ref{tab:polePos}, there are 
significant
differences in the pole positions for 
three
nucleon resonances. Table \ref{tab:photonDecayAmp} displays the differences in the photo-decay amplitudes for these three resonances. 

The new data have a noticeable impact on the partial-wave content of the $\gamma p\to K^0\Sigma^+$ channel. In JuBo2022 the dominant contribution came from the $P_{13}$ partial wave, followed by $P_{33}$. This order is now reversed. Accordingly, we observe major changes in the  $E_{1+}$ and $M_{1+}$ multipole amplitudes (see Figure~\ref{fig:E1+M1+JuBo}) and also the pole position of the $N(1900)3/2^+$ changes from $1905- i46.5$~MeV~\cite{Ronchen:2022hqk} to $1893-i 52$~MeV. 

Another remarkable shift concerns the $\Delta(1910)1/2^+$ resonance, which was very broad ($\sim 550$~MeV) but is now narrower by almost $120$~MeV (see Table \ref{tab:polePos}). In addition the real part of the pole moved  by almost 60~MeV to a lower value. Moreover, the contribution of the $P_{31}$ partial wave to the $K^0\Sigma^+$ total cross section is increased. These observations emphasize the importance of polarization observables to determine the amplitude and to disentangle the isospin content in the $K\Sigma$ channels. Further differences in pole positions occur for states in higher partial waves where the resonance parameters are less stable in general. However, it is noteworthy that the width of the $N(2190)7/2^-$ was much broader in previous JuBo fits than in the new fit.


\clearpage

\onecolumngrid

\noindent \begin{center}
\begin{figure}[h]
\noindent \centering{}\includegraphics[width=0.9\columnwidth]{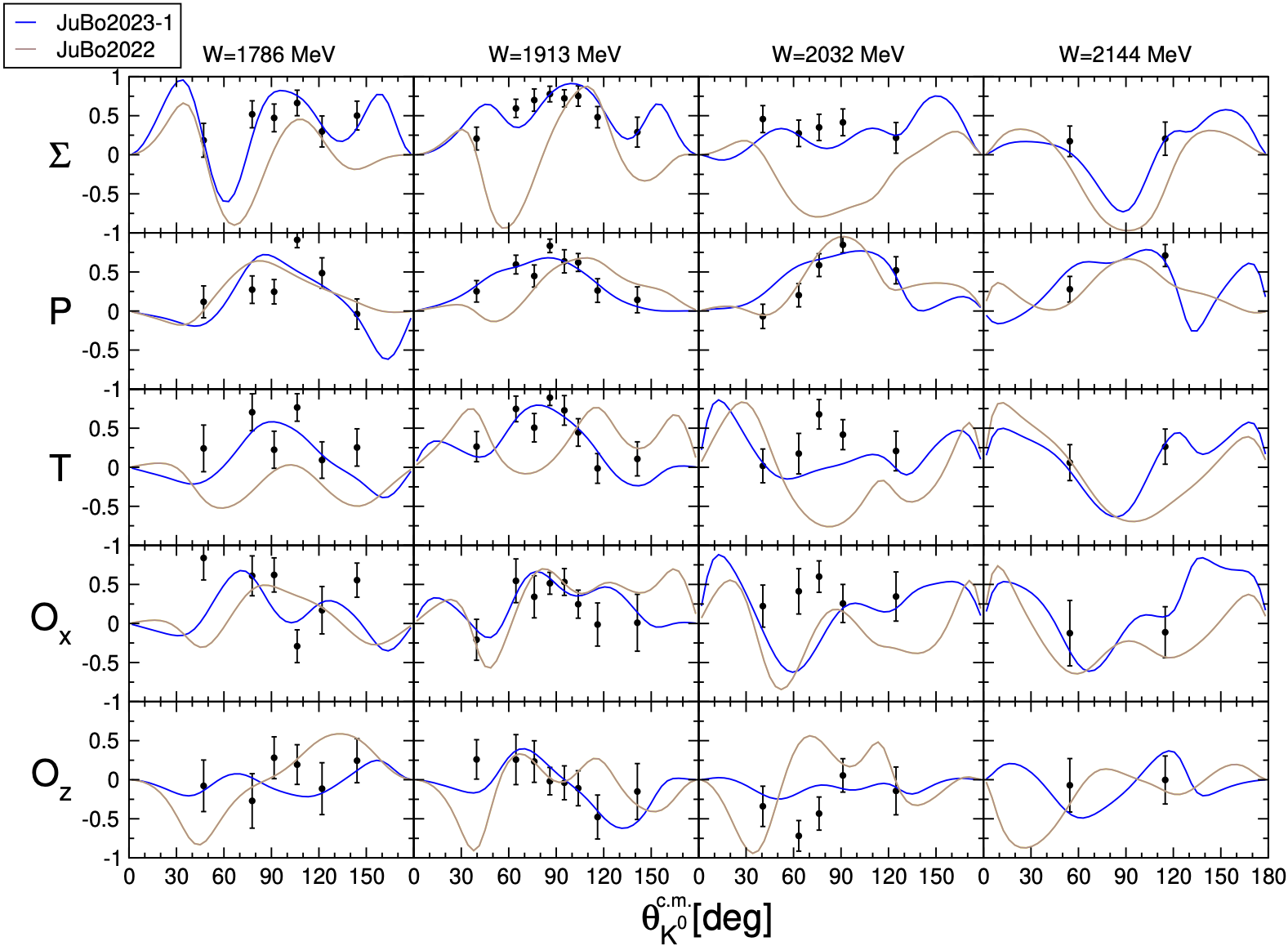}\caption{\label{fig:CompJuBo}Comparison of results against the Juelich-Bonn dynamical coupled channels model ~\cite{Ronchen:2012eg}. Data points from this work are in black, the brown line is the 2022 model prediction, and the blue line is after including the data. The dependence on $\theta^{c.m.}_{K^0}$ is shown for each observable for four centre-of-mass energy bins.}
\end{figure}
\par\end{center}
\twocolumngrid

\noindent \begin{center}
\begin{figure}[h]
\noindent \centering{}\includegraphics[width=0.9\columnwidth]{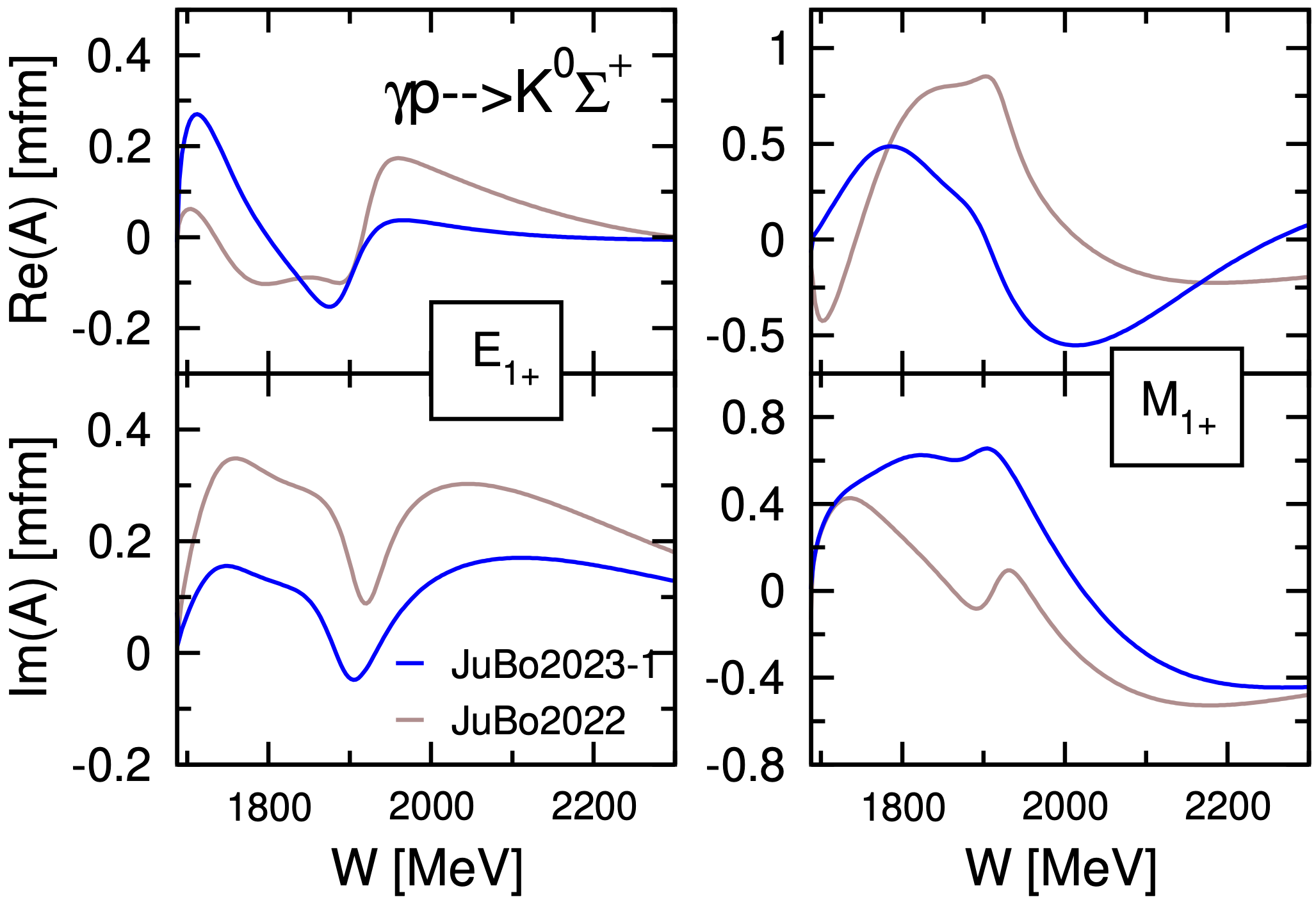}
\caption{\label{fig:E1+M1+JuBo} Real (upper row) and imaginary parts (lower row) of the $E_{1+}$ (left) and $M_{1+}$ (right) multipoles for the reaction $\gamma p\to K^0\Sigma^+$ from the Juelich-Bonn dynamical coupled channels model. Brown lines: JuBo2022 solution. Blue lines: JuBo2023-1 solution that includes the data of this work in the fits.}
\end{figure}
\par\end{center}

\begin{table*}[h]
\centering{}%
\begin{tabular}{|c|c|c|}
\hline 
 & \multicolumn{2}{c|}{Reduced $\chi^2$}\tabularnewline
\hline 
 & JuBo2022 before fit & JuBo2023-1 after fit\tabularnewline
\hline 
\hline 
$\Sigma$ & 13.35 & 1.01\tabularnewline
\hline 
$P$ & 2.43 & 1.44\tabularnewline
\hline 
$T$ & 8.97 & 1.34\tabularnewline
\hline 
$O_x$ & 2.94 & 1.96\tabularnewline
\hline 
$O_z$ & 3.30 & 0.91\tabularnewline
\hline 
\end{tabular}\caption[Reduced $\chi^2$ values before and after the inclusion of the new data.]{Reduced $\chi^2$ values before (JuBo2022 solution~\cite{Ronchen:2022hqk}) and after the inclusion of the new data (JuBo2023-1 solution).}
\label{tab:fitChiSq}
\end{table*}

\begin{table*}[h]
\centering{}%
\begin{tabular}{|c|c|c|c|c|}
\hline 
 & \multicolumn{2}{c|}{Pole position (MeV)} & \multicolumn{2}{c|}{Width (MeV)}\tabularnewline 
\hline 
\hline 
 & Before fitting & After fitting & Before fitting & After fitting\tabularnewline
\hline 
$\Delta(1910)$ $J^P=1/2^+$& 1802 (11) & 1745 & 550 (22) & 433 \tabularnewline
\hline
$N(1900)$  $J^P=3/2^+$ & 1905 (3) & 1893 & 93 (4) & 105\tabularnewline
\hline 

$N(2190)$  $J^P=7/2^-$ & 1965 (12) & 1946 & 288 (66) & 162 \tabularnewline
\hline 
\end{tabular}\caption[Pole positions of resonances affected by the new data, before  and after fitting.]{Pole positions of resonances affected by the new data, before (JuBo2022 solution~\cite{Ronchen:2022hqk}) and after fitting (JuBo2023-1 solution).}
\label{tab:polePos}
\end{table*}

\begin{table*}[h]
\centering{}%
\begin{tabular}{|cr|c|c|c|c|}
\hline 
 && \multicolumn{2}{c|}{Modulus (GeV$^{-1/2}$)} & \multicolumn{2}{c|}{Phase}\tabularnewline 
\hline 
\hline 
& & Before fitting & After fitting & Before fitting & After fitting \tabularnewline
\hline 
$\Delta(1910)$ $J^P=1/2^+$ & $A_{1/2}$ & -0.446 (0.072) & -0.449  &  -69.50$^\circ$ (21) &  -157.28$^\circ$\tabularnewline
\hline 
$N(1900)$  $J^P=3/2^+$ & $A_{1/2}$& 0.0091 (0.0027) & 0.0077 & 80.45$^\circ$ (23) & 26.98$^\circ$\tabularnewline
&$A_{3/2}$ & -0.0077 (0.0034) &  -0.029 & -42.38$^\circ$ (23) &-30.1$^\circ$   \tabularnewline
\hline 
$N(2190)$  $J^P=7/2^-$ &$A_{1/2}$& 0.015 (0.008) &  0.005 &  -68.92$^\circ$ (17) & -174.10$^\circ$\tabularnewline
&$A_{3/2}$& -0.062 (0.022)   & -0.031  & -0.54$^\circ$ (26)& -58.17$^\circ$\tabularnewline
\hline 
\end{tabular}\caption[Photon decay amplitudes affected by the new data, before and after fitting.]{Photon decay amplitudes affected by the new data, before (JuBo2022 solution~\cite{Ronchen:2022hqk}) and after fitting (JuBo2023-1 solution).}.
\label{tab:photonDecayAmp}
\end{table*}


\section{\label{sec:conclusions}Conclusions}

Five polarization observables $(\obs{\Sigma}, \obs{O_x}, \obs{O_z}, \obs{T}, \obs{P})$ have been extracted for the reaction  $\vec{\gamma} p \rightarrow K_{S}^0 \Sigma^{+}$ in the nucleon resonance region using a linearly polarized photon beam and the CLAS detector. Comparison with state-of-the-art coupled-channels calculations demonstrate that even though the measurements were made at relatively few kinematic points, nonetheless they carry sufficient information to be sensitive to the details of several light baryon resonances and are thus an important addition to the database of photoproduction results.



\begin{acknowledgments}
The authors gratefully acknowledge the work of Jefferson Lab staff in the Accelerator and Physics Divisions.
This work was supported by: the United Kingdom's Science and Technology Facilities Council (STFC) from grant numbers ST/V00106X/1 and ST/P004458/1;
the Chilean Comisi\'on Nacional de Investigaci\'on Cient\'ifica y Tecnol\'ogica (CONICYT);
the Italian Istituto Nazionale di Fisica Nucleare;
the French Centre National de la Recherche Scientifique;
the French Commissariat \`{a} l'Energie Atomique;
the U.S.~National Science Foundation;
the National Research Foundation of Korea (NRF).
Jefferson Science Associates, LLC, operates the Thomas Jefferson National Accelerator Facility for the the U.S.~Department of Energy under contract DE-AC05-06OR23177. 
The authors gratefully acknowledge computing time on the supercomputer JURECA at Forschungszentrum Jülich under grant no. ``baryonspectro". This work was supported in part by the Deutsche Forschungsgemeinschaft (DFG) from Project-ID 196253076-TRR 110. 
\end{acknowledgments}





\bibliographystyle{h-physrev}
\bibliography{g8_KY_references}


\end{document}

%% file: authors.tex

\newcommand*{\GLASGOW}{University of Glasgow, Glasgow G12 8QQ, United Kingdom}
\newcommand*{\GLASGOWindex}{1}
\affiliation{\GLASGOW}

\newcommand*{\JULICH}{Institute for Advanced Simulation and J\"ulich Center for Hadron Physics, Forschungszentrum J\"ulich, 52425 J\"ulich, Germany}
\newcommand*{\JULICHindex}{2}
\affiliation{\JULICH}

\newcommand*{\ANL}{Argonne National Laboratory, Argonne, Illinois 60439}
\newcommand*{\ANLindex}{1}
\affiliation{\ANL}

\newcommand*{\ASU}{Arizona State University, Tempe, Arizona 85287-1504}
\newcommand*{\ASUindex}{2}
\affiliation{\ASU}
\newcommand*{\CMU}{Carnegie Mellon University, Pittsburgh, Pennsylvania 15213}
\newcommand*{\CMUindex}{3}
\affiliation{\CMU}
\newcommand*{\CUA}{Catholic University of America, Washington, D.C. 20064}
\newcommand*{\CUAindex}{4}
\affiliation{\CUA}
\newcommand*{\SACLAY}{IRFU, CEA, Universit\'{e} Paris-Saclay, F-91191 Gif-sur-Yvette, France}
\newcommand*{\SACLAYindex}{5}
\affiliation{\SACLAY}
\newcommand*{\CNU}{Christopher Newport University, Newport News, Virginia 23606}
\newcommand*{\CNUindex}{6}
\affiliation{\CNU}
\newcommand*{\UCONN}{University of Connecticut, Storrs, Connecticut 06269}
\newcommand*{\UCONNindex}{7}
\affiliation{\UCONN}
\newcommand*{\DUKE}{Duke University, Durham, North Carolina 27708-0305}
\newcommand*{\DUKEindex}{8}
\affiliation{\DUKE}
\newcommand*{\DUQUESNE}{Duquesne University, 600 Forbes Avenue, Pittsburgh, PA 15282 }
\newcommand*{\DUQUESNEindex}{9}
\affiliation{\DUQUESNE}
\newcommand*{\FU}{Fairfield University, Fairfield CT 06824}
\newcommand*{\FUindex}{10}
\affiliation{\FU}
\newcommand*{\FERRARAU}{Universita' di Ferrara , 44121 Ferrara, Italy}
\newcommand*{\FERRARAUindex}{11}
\affiliation{\FERRARAU}
\newcommand*{\FIU}{Florida International University, Miami, Florida 33199}
\newcommand*{\FIUindex}{12}
\affiliation{\FIU}
\newcommand*{\FSU}{Florida State University, Tallahassee, Florida 32306}
\newcommand*{\FSUindex}{13}
\affiliation{\FSU}
\newcommand*{\GWUI}{The George Washington University, Washington, DC 20052}
\newcommand*{\GWUIindex}{14}
\affiliation{\GWUI}
\newcommand*{\GSIFFN}{GSI Helmholtzzentrum fur Schwerionenforschung GmbH, D-64291 Darmstadt, Germany}
\newcommand*{\GSIFFNindex}{15}
\affiliation{\GSIFFN}
\newcommand*{\INFNFE}{INFN, Sezione di Ferrara, 44100 Ferrara, Italy}
\newcommand*{\INFNFEindex}{16}
\affiliation{\INFNFE}
\newcommand*{\INFNGE}{INFN, Sezione di Genova, 16146 Genova, Italy}
\newcommand*{\INFNGEindex}{17}
\affiliation{\INFNGE}
\newcommand*{\INFNRO}{INFN, Sezione di Roma Tor Vergata, 00133 Rome, Italy}
\newcommand*{\INFNROindex}{18}
\affiliation{\INFNRO}
\newcommand*{\INFNTUR}{INFN, Sezione di Torino, 10125 Torino, Italy}
\newcommand*{\INFNTURindex}{19}
\affiliation{\INFNTUR}
\newcommand*{\INFNPAV}{INFN, Sezione di Pavia, 27100 Pavia, Italy}
\newcommand*{\INFNPAVindex}{20}
\affiliation{\INFNPAV}
\newcommand*{\ORSAY}{Universit'{e} Paris-Saclay, CNRS/IN2P3, IJCLab, 91405 Orsay, France}
\newcommand*{\ORSAYindex}{21}
\affiliation{\ORSAY}
\newcommand*{\JMU}{James Madison University, Harrisonburg, Virginia 22807}
\newcommand*{\JMUindex}{22}
\affiliation{\JMU}
\newcommand*{\KNU}{Kyungpook National University, Daegu 41566, Republic of Korea}
\newcommand*{\KNUindex}{23}
\affiliation{\KNU}
\newcommand*{\LAMAR}{Lamar University, 4400 MLK Blvd, PO Box 10046, Beaumont, Texas 77710}
\newcommand*{\LAMARindex}{47}
\affiliation{\LAMAR}
\newcommand*{\MIT}{Massachusetts Institute of Technology, Cambridge, Massachusetts  02139-4307}
\newcommand*{\MITindex}{24}
\affiliation{\MIT}
\newcommand*{\MISS}{Mississippi State University, Mississippi State, MS 39762-5167}
\newcommand*{\MISSindex}{25}
\affiliation{\MISS}
\newcommand*{\UNH}{University of New Hampshire, Durham, New Hampshire 03824-3568}
\newcommand*{\UNHindex}{26}
\affiliation{\UNH}
\newcommand*{\NMSU}{New Mexico State University, PO Box 30001, Las Cruces, NM 88003, USA}
\newcommand*{\NMSUindex}{27}
\affiliation{\NMSU}
\newcommand*{\NSU}{Norfolk State University, Norfolk, Virginia 23504}
\newcommand*{\NSUindex}{28}
\affiliation{\NSU}
\newcommand*{\OHIOU}{Ohio University, Athens, Ohio  45701}
\newcommand*{\OHIOUindex}{29}
\affiliation{\OHIOU}
\newcommand*{\ODU}{Old Dominion University, Norfolk, Virginia 23529}
\newcommand*{\ODUindex}{30}
\affiliation{\ODU}
\newcommand*{\JLUGiessen}{II Physikalisches Institut der Universitaet Giessen, 35392 Giessen, Germany}
\newcommand*{\JLUGiessenindex}{31}
\affiliation{\JLUGiessen}
\newcommand*{\RPI}{Rensselaer Polytechnic Institute, Troy, New York 12180-3590}
\newcommand*{\RPIindex}{32}
\affiliation{\RPI}
\newcommand*{\ROMAII}{Universita' di Roma Tor Vergata, 00133 Rome Italy}
\newcommand*{\ROMAIIindex}{33}
\affiliation{\ROMAII}
\newcommand*{\MSU}{Skobeltsyn Institute of Nuclear Physics, Lomonosov Moscow State University, 119234 Moscow, Russia}
\newcommand*{\MSUindex}{34}
\affiliation{\MSU}
\newcommand*{\SCAROLINA}{University of South Carolina, Columbia, South Carolina 29208}
\newcommand*{\SCAROLINAindex}{35}
\affiliation{\SCAROLINA}
\newcommand*{\TEMPLE}{Temple University,  Philadelphia, PA 19122 }
\newcommand*{\TEMPLEindex}{36}
\affiliation{\TEMPLE}
\newcommand*{\JLAB}{Thomas Jefferson National Accelerator Facility, Newport News, Virginia 23606}
\newcommand*{\JLABindex}{37}
\affiliation{\JLAB}
\newcommand*{\UTFSM}{Universidad T\'{e}cnica Federico Santa Mar\'{i}a, Casilla 110-V Valpara\'{i}so, Chile}
\newcommand*{\UTFSMindex}{38}
\affiliation{\UTFSM}
\newcommand*{\BRESCIA}{Universit`{a} degli Studi di Brescia, 25123 Brescia, Italy}
\newcommand*{\BRESCIAindex}{39}
\affiliation{\BRESCIA}
\newcommand*{\UCR}{University of California Riverside, 900 University Avenue, Riverside, CA 92521, USA}
\newcommand*{\UCRindex}{40}
\affiliation{\UCR}
\newcommand*{\URICH}{University of Richmond, Richmond, Virginia 23173}
\newcommand*{\URICHindex}{41}
\affiliation{\URICH}
\newcommand*{\YORK}{University of York, York YO10 5DD, United Kingdom}
\newcommand*{\YORKindex}{42}
\affiliation{\YORK}
\newcommand*{\VT}{Virginia Tech, Blacksburg, Virginia   24061-0435}
\newcommand*{\VTindex}{43}
\affiliation{\VT}
\newcommand*{\VIRGINIA}{University of Virginia, Charlottesville, Virginia 22901}
\newcommand*{\VIRGINIAindex}{44}
\affiliation{\VIRGINIA}
\newcommand*{\WM}{College of William and Mary, Williamsburg, Virginia 23187-8795}
\newcommand*{\WMindex}{45}
\affiliation{\WM}
\newcommand*{\YEREVAN}{Yerevan Physics Institute, 375036 Yerevan, Armenia}
\newcommand*{\YEREVANindex}{46}
\affiliation{\YEREVAN}

\newcommand*{\NOWJLAB}{Thomas Jefferson National Accelerator Facility, Newport News, Virginia 23606}
\newcommand*{\NOWISU}{Idaho State University, Pocatello, Idaho 83209}
\newcommand*{\NOWKSU}{King Saud University, Riyadh, Saudi Arabia}


\author{L.~Clark}
\altaffiliation[Current address: ]{Information Services, Durham University, U.K.}
\affiliation{\GLASGOW}

\author{{B.~McKinnon}
\orcidlink{0000-0002-5550-0980}}
\email[Corresponding author: ]{Bryan.McKinnon@glasgow.ac.uk}
\affiliation{\GLASGOW}

\author{{D.G.~Ireland}
\orcidlink{0000-0001-7713-7011}}
\affiliation{\GLASGOW}

\author{{D.I.~Glazier}
\orcidlink{0000-0002-8929-6332}}
\affiliation{\GLASGOW}

\author{{K.~Livingston}
\orcidlink{0000-0001-7166-7548}}
\affiliation{\GLASGOW}

\author{{D.~R\"onchen}
\orcidlink{0000-0002-9650-8401}}
\affiliation{\JULICH}

\author {A.G.~Acar}
\affiliation{\YORK}
\author {P.~Achenbach} 
\affiliation{\JLAB}
\author {J.S.~Alvarado} 
\affiliation{\ORSAY}
\author {M.J.~Amaryan} 
\affiliation{\ODU}
% ...

\collaboration{The CLAS Collaboration}
\noaffiliation

